\documentclass[aps,prc,preprint,amsmath,amssymb,showpacs,superscriptaddress]{revtex4}

\usepackage{graphicx}
\usepackage{dcolumn}
\usepackage{bm}
\usepackage{longtable}
\usepackage{color}
\usepackage{CJK}

\begin{document}

\title{Crucial test for covariant density functional theory with new and accurate mass measurements from Sn to Pa}

\author{P. W. Zhao}%
\affiliation{State Key Laboratory of Nuclear Physics and Technology, School of Physics, Peking University, Beijing 100871, China}
\author{L. S. Song}%
\affiliation{State Key Laboratory of Nuclear Physics and Technology, School of Physics, Peking University, Beijing 100871, China}
\author{B. Sun}%
\affiliation{School of Physics and Nuclear Energy Engineering, Beihang University, Beijing 100191, China}
\affiliation{Justus-Liebig-Universit\"{a}t Giessen, Heinrich-Buff-Ring 14, Giessen 35392, Germany}
\author{H. Geissel}%
\affiliation{Justus-Liebig-Universit\"{a}t Giessen, Heinrich-Buff-Ring 14, Giessen 35392, Germany}
\affiliation{Gesellschaft f\"{u}r Schwerionenforschung (GSI), D-64291 Darmstadt, Germany}
\author{J. Meng}%
\affiliation{State Key Laboratory of Nuclear Physics and Technology, School of Physics, Peking University, Beijing 100871, China}
\affiliation{School of Physics and Nuclear Energy Engineering, Beihang University, Beijing 100191, China}
\affiliation{Department of Physics, University of Stellenbosch, Stellenbosch, South Africa}

\begin{abstract}
  The covariant density functional theory with the point-coupling interaction PC-PK1 is compared with new and accurate experimental masses in the element range from 50 to 91. The experimental data are from a mass measurement performed with the storage ring mass spectrometry at Gesellschaft f\"{u}r Schwerionenforschung (GSI) [Chen \emph{et al.}, Nucl. Phys. A \textbf{882} 71 (2012)] . Although the microscopic theory contains only 11 parameters, it agrees well with the experimental data. The comparison is characterized by a rms deviation of 0.859 MeV. For even-even nuclei, the theory agrees within about 600 keV. Larger deviations are observed in this comparison for the odd-A and odd-odd nuclei. Improvements and possible reasons for the deviations are discussed in this contribution as well.
\end{abstract}


\maketitle

Nuclear masses play a vital role not only in the nuclear physics but also in weak-interaction studies and astrophysics. In particular, the mass of heavy neutron-rich nuclei is one of the basic topics in nuclear physics and is essential for understanding the rapid neutron-capture nucleosynthesis process (\emph{r} process).

During recent decades, great achievements in mass measurements of neutron-rich nuclei have been made, thanks to the applications of cyclotron, storage ring, and penning trap facilities~\cite{Lunney2003Rev.Mod.Phys.1021}. Despite the experimental progress, the masses of a large number of neutron-rich nuclei relevant to the \emph{r} process remain unmeasured due to difficulties in production, separation, and detection. Therefore, reliable theoretical predictions of the nuclear mass are imperative at the present time.

The theoretical determination of nuclear masses can be traced to the von Weizs\"{a}cker mass formula, which was proposed based on the famous liquid drop model (LDM)~\cite{Weizsacker1935Z.Physik431}. Since then, tremendous effort has been made in pursuing different possible extensions of the LDM which are known as the macroscopic-microscopic models, for example, the finite-range droplet model (FRDM)~\cite{Moller1995At.DataNucl.DataTables185}, the extended Thomas-Fermi plus Strutinsky integral with shell quenching (ETFSI-Q) model~\cite{Pearson1996Phys.Lett.B455}, and the Weizs\"{a}cker-Skyrme (WS) model~\cite{Wang2010Phys.Rev.C44304}. It turns out that these macroscopic-microscopic models work pretty well in the description of known nuclides, but their predictions show a large deviation for very neutron-rich nuclides. In addition, the local mass relations such as the isobaric multiplet mass equation (IMME)~\cite{Ormand1997Phys.Rev.C2407}, the Garvey-Kelson (GK) relations~\cite{Barea2008Phys.Rev.C41304}, and the residual proton-neutron interactions~\cite{Fu2011Phys.Rev.C34311} are also used to give predictions of unmeasured masses.

In principle, an ideal mass formula would be one in which the masses of all nuclei are derived from the basic nucleonic interactions. In this regard, the microscopic-rooted mass model, which treats the macroscopic part and the microscopic corrections in a unified framework, is usually believed to have a more reliable extrapolation to the unknown regions.
In the past decade, a series of microscopic-rooted mass models based on the nonrelativistic density functional theory (DFT) have been developed with the Hartree-Fock-Bogoliubov (HFB) method and have achieved great success in the description of known masses (see Refs.~\cite{Goriely2009Phys.Rev.Lett.152503,Goriely2009Phys.Rev.Lett.242501,Goriely2010Phys.Rev.C35804} and references therein). In these models, the model parameters are fitted to essentially all the experimental mass data.

Apart from the nonrelativistic DFT, the covariant density functional theory (CDFT) has also received wide attention due to its successful description of many nuclear phenomena~\cite{Ring1996Prog.Part.Nucl.Phys.193,Meng2006Prog.Part.Nucl.Phys.470,Vretenar2005Phys.Rep.101,Niksic2011Prog.Part.Nucl.Phys.519}.
There exist a number of attractive features in the CDFT, especially in its practical applications in the self-consistent relativistic mean-field (RMF) framework. The most obvious one is the natural inclusion of the spin degree of freedom, and the relativistic effects are responsible for the existence of the approximate pseudospin
symmetry~\cite{Ginocchio2005Phys.Rep.165,Meng1999Phys.Rev.C154,Meng1998Phys.Rev.C628,Liang2011Phys.Rev.C41301,Guo2012Phys.Rev.C21302,Lu2012Phys.Rev.Lett.72501} in the nuclear single-particle spectra and spin symmetry in the antinucleon spectra~\cite{Zhou2003Phys.Rev.Lett.262501}.
Moreover, it is of particular importance that the CDFT includes nuclear magnetism~\cite{Koepf1989Nucl.Phys.A61}, which plays an important role in the microscopic description of the nuclear magnetic moments~\cite{Yao2006Phys.Rev.C24307,Arima2011SciChinaSerG-PhysMechAstron188,Li2011Sci.ChinaPhys.Mech.Astron.204,Li2011Prog.Theor.Phys.1185,Wei2012Prog.Theor.Phys.400} and nuclear rotations~\cite{Afanasjev2000Nucl.Phys.A196,Zhao2012Phys.Rev.C54310,Zhao2011Phys.Rev.Lett.122501,Zhao2011Phys.Lett.B181}.

The first RMF mass table was reported in Ref.~\cite{Hirata1997Nucl.Phys.A438} for about 2000 even-even nuclei with $8 \le Z\le 120$ up to the proton and neutron drip lines but without including pairing correlations. Later on, the ground-state properties of 1315 even-even nuclei with $10 \le Z\le 98$ were calculated by including the pairing correlations with BCS method~\cite{Lalazissis1999At.DataNucl.DataTables1}. In Ref.~\cite{Geng2005Prog.Theor.Phys.785}, by using a state-dependent BCS method with zero-range $\delta$ force, the first systematic study of the ground-state properties of more than 7000 nuclei ranging from the proton drip line to the neutron drip line was performed with the meson-exchange effective interaction TMA. This mass table works well in the \emph{r}-process nucleosynthesis calculations~\cite{Sun2008Phys.Rev.C025806,Niu2009Phys.Rev.C065806,Li2012ActaPhys.Sin72601,Xu2012}.

Very recently, a new point-coupling effective interaction PC-PK1 has been proposed by fitting to observables of 60 selected spherical nuclei, including the binding energies, charge radii, and empirical pairing gaps~\cite{Zhao2010Phys.Rev.C54319}. This effective interaction particularly improves the description for isospin dependence of binding energies and it has been successfully used in describing the Coulomb displacement energies between mirror nuclei~\cite{Sun2011Sci.ChinaSer.G-Phys.Mech.Astron.210}, fission barriers~\cite{Lu2012Phys.Rev.C11301}, nuclear rotations~\cite{Zhao2012Phys.Rev.C54310,Zhao2011Phys.Rev.Lett.122501,Zhao2011Phys.Lett.B181}, etc.
On the other hand, a new opportunity has been opened for a crucial test of the predictive power of the theories with the comparison to accurate new mass values of 53 heavy neutron-rich isotopes from Sn to Pa measured with the storage ring mass spectrometry at GSI~\cite{Chen2012Nucl.Phys.A71}. This mass measurement is characterized by a small systematic error of about 10 keV, which is valid for all data. The overall mean experimental accuracy is about 19 keV. From these 53 values, 31 have been measured for the first time, whereas for the additional 22 nuclides the experimental error has been significantly improved. Since none of these data have been used to determine the parameters of the CDFT with PC-PK1 and most of them have not been used in the fitting of the widely used mass models in the market, these data provide a very good test for the mass-prediction power for nuclei in this region. The work for all the measured masses with CDFT definitely should be done in the future.

In this work, the CDFT with the point-coupling interaction PC-PK1 is applied to investigate the new experimental masses of 53 heavy neutron-rich isotopes at the storage ring mass spectrometry at GSI, including 31 cases measured for the first time~\cite{Chen2012Nucl.Phys.A71}. The theoretical predictions are compared with the experimental data as well as other theoretical results.

The starting point of the CDFT is a general effective Lagrangian density where the nucleons are coupled with zero-range point-coupling interaction~\cite{Nikolaus1992Phys.Rev.C1757,Burvenich2002Phys.Rev.C44308,Zhao2010Phys.Rev.C54319}. By means of the mean-field approximation and the no-sea approximation, one could obtain the corresponding Kohn-Sham equation, which has the form of a Dirac equation
\begin{equation}
 \label{DiracEq}
   [\bm{\alpha}\cdot\bm{p}+\beta(m+S)+V]\psi_k=\epsilon_k\psi_k,
\end{equation}
with the scalar $S(\bm{r})$ and vector $V(\bm{r})$ potentials,
\begin{subequations}
\begin{eqnarray}
 S(\bm{r})          &=&\alpha_S\rho_S+\beta_S\rho^2_S+\gamma_S\rho^3_S+\delta_S\triangle\rho_S,\\
 V(\bm{r})          &=&\alpha_V\rho_V +\gamma_V \rho_V^3
                      +\delta_V\triangle \rho_V + e A_0\nonumber\\
 &&+\alpha_{TV}\tau_3 \rho_{TV}+\delta_{TV}\triangle\tau_3\rho_{TV}.
\end{eqnarray}
\end{subequations}
Here, $m$ is the nucleon mass, and $\alpha_S$, $\alpha_V$, $\alpha_{TV}$, $\beta_S$, $\gamma_S$, $\gamma_V$, $\delta_S$, $\delta_V$, and $\delta_{TV}$ are the coupling constants. The Coulomb field $A_0$ is determined by Poisson's equation.

The iterative solution of these equations yields the single-particle energies and expectation values of total energy, quadrupole moments, etc. Here, we present only the expectation value of total energy
\begin{eqnarray} \label{Energy}
 E_{\rm DF}&=& \int d^3r~\left\{\sum_k\,v_k^2~{\psi^\dagger_k (\bm{r})
    \left(\bm{\alpha}\cdot\bm{p} + \beta m\right )\psi_k(\bm{r})}  \right.\nonumber\\
 &&+ \frac{\alpha_S}{2}\rho_S^2+\frac{\beta_S}{3}\rho_S^3
    + \frac{\gamma_S}{4}\rho_S^4+\frac{\delta_S}{2}\rho_S\triangle\rho_S\nonumber \\
 && + \frac{\alpha_V}{2}\rho_V^2+ \frac{\gamma_V}{4}\rho_V^4 +
       \frac{\delta_V}{2}\rho_V\triangle \rho_V \nonumber\\
 &&\left.+ \frac{\alpha_{TV}}{2}\rho^2_{TV}+\frac{\delta_{TV}}{2} \rho_{TV}\triangle \rho_{TV}+  \frac{1}{2}eA_0 \rho_V^p\right\},
\end{eqnarray}
where $v_k^2$ is the occupation probabilities of single-particle state and $\rho_S(\bm{r})$, $\rho_V(\bm{r})$, and $\rho_{TV}(\bm{r})$ are the local densities in scalar, vector, and isovector-vector channels, respectively.

For open-shell nuclei, pairing correlations are taken into account by the BCS method with a zero-range $\delta$ force. Thus, we have to add to the functional Eq.~(\ref{Energy}) a pairing energy term,
\begin{equation}
\label{PairingDFT}
E_{\rm pair}=
-\sum_{\tau=n,p} \dfrac{V_\tau}{4}\int d^3r\kappa^\ast_\tau(\bm{r}) \kappa_\tau(\bm{r}),
\end{equation}
where $V_\tau$ is the constant pairing strength, and the
pairing tensor $\kappa(\bm{r})$ reads
\begin{equation}
 \kappa(\bm{r})
 =-2\sum_{k>0}f_ku_kv_k\vert\psi_k(\bm{r})\vert^2
\end{equation}
with $f_k$ being a smooth cutoff factor~\cite{Krieger1990Nucl.Phys.A275,Bender2000Eur.Phys.J.A59}.

As the translational symmetry is broken in the mean-field approximation, the center-of-mass (c.m.) correction should be made for the spurious c.m. motion. Here, we adopt the microscopic c.m. correction~\cite{Bender2000Eur.Phys.J.A467,Zhao2009Chin.Phys.Lett.112102}
\begin{equation}
 E_{\rm c.m.}=-\frac{1}{2mA}\langle\hat{\bm P}^{2}_{\rm c.m.}\rangle,
\end{equation}
with $A$ being the mass number and $\hat{\bm{P}}_{\rm c.m.}=\sum_i^A \hat{\bm{p}}_i$ being the total momentum in the c.m. frame.

Similarly, the rotational symmetry is also violated for the deformed nuclei. Therefore, we further introduce the correction energy
\begin{equation}\label{Erot}
E_{\rm rot}=-\frac{\hbar^2}{2\cal I}\langle \hat{J}^2\rangle
\end{equation}
with $\hat{J}$ the angular momentum operator and $\cal I$ the moment of inertia calculated from the Inglis-Belyaev formula~\cite{Inglis1956Phys.Rev.1786,Belyaev1961Nucl.Phys.A322,Volkov1972Phys.Lett.B1}.

Finally, the total energy for the nuclear system reads
\begin{equation}
E= E_{\rm DF}+E_{\rm pair}+E_{\rm c.m.}+E_{\rm rot},
\end{equation}
and the corresponding binding energy is $E_{\rm B} = -E$.

In this work, the Dirac equation is solved on the basis of an axially deformed harmonic oscillator potential~\cite{Ring1997Comput.Phys.Commun.77}. A basis of 16 major oscillator shells is used in the calculations, and convergence has been tested in the calculations with 18 major shells.

In Table~\ref{Tabel}, the calculated binding energies $E^{\rm Cal}_{\rm B}$, the rotational correction energies (RCEs) $E^{\rm rot}_{\rm B}=-E_{\rm rot}$, and the quadrupole deformations $\beta_2$ for nuclei with masses measured at the storage ring mass spectrometry at GSI~\cite{Chen2012Nucl.Phys.A71} are listed together with the differences between the data $E^{\rm Exp}_{\rm B}$ and the calculated binding energies $\Delta E=E^{\rm Exp}_{\rm B}-E^{\rm Cal}_{\rm B}$.

It is found that most nuclei are deformed except the nuclei with proton number close to the magic numbers of 50 and 82.
Note that it is not necessary to make rotational correction to the binding energy for a spherical nucleus, since it preserves the rotational symmetry. Therefore, we here consider the RCEs only for the deformed nuclei with $|\beta_2|> 0.02$. The present calculations with PC-PK1 reproduce the experimental data very well and the root mean square (rms) deviation results in 0.859 MeV for 53 nuclei, and 0.805 MeV for the 31 cases measured for the first time.

\setlength{\LTcapwidth}{3.5in}
\begin{longtable}{@{\extracolsep{0.08in}}cccccccc}
\caption{The experimental binding energies $E^{\rm Exp}_{\rm B}$, calculated binding energies $E^{\rm Cal}_{\rm B}$, rotational correction energies $E^{\rm rot}_{\rm B}$, and quadrupole deformations $\beta_2$ with the PC-PK1, as well as the differences between the data and the calculated results $\Delta E$. The root mean square (rms) deviation $\Delta$ is listed in the last row. The energies are shown in million
electron volts (MeV).}
\label{Tabel}\\
\hline\hline
\noalign{\vspace{3pt}}
Element  &    Z & A   & $E^{\rm Exp}_{\rm B}$  &    $E^{\rm Cal}_{\rm B}$    &    $E^{\rm rot}_{\rm B}$    &   $\beta_2$  & $\Delta E$\\
\noalign{\vspace{3pt}}
\hline
\noalign{\vspace{3pt}}
\endfirsthead
\caption{(Continued).}\\%
\hline\hline
\noalign{\vspace{3pt}}
Element  &    Z & A   & $E^{\rm Exp}_{\rm B}$  &    $E^{\rm Cal}_{\rm B}$   &    $E^{\rm rot}_{\rm B}$  &   $\beta_2$  & $\Delta E$\\
\noalign{\vspace{3pt}}
\hline
\noalign{\vspace{3pt}}
\endhead
\hline
\endfoot \hline\hline
\endlastfoot
Sn       &  50  & 128 &   1077.20      &1078.67 &0.00 & 0.00  &-1.47       \\
Sb       &  51  & 133 &   1112.28      &1113.70 &0.00 & -0.02 &-1.42      \\
Te       &  52  & 136 &   1131.28      &1131.48 &1.15 & 0.09  &-0.20      \\
La       &  57  & 144 &   1192.29      &1191.36 &0.71 & 0.19  &0.93      \\
Ce       &  58  & 146 &   1208.38      &1208.25 &2.00 & 0.20  &0.13      \\
Pt       &  78  & 202 &   1591.54      &1590.09 &0.00 & 0.00  &1.45      \\
Au       &  79  & 202 &   1592.40      &1590.78 &0.57 & -0.08 &1.62      \\
Hg       &  80  & 207 &   1624.09      &1624.17 &0.00 & -0.02 &-0.08      \\
Tl       &  81  & 213 &   1653.45      &1652.30 &0.00 & 0.00  &1.15      \\
Bi       &  83  & 217 &   1677.18      &1677.10 &0.87 & -0.04 &0.08      \\
Bi       &  83  & 218 &   1680.77      &1679.28 &0.04 & 0.07  &1.49      \\
Po       &  84  & 219 &   1688.58      &1687.34 &0.33 & 0.11  &1.24      \\
Po       &  84  & 220 &   1694.07      &1694.04 &1.44 & 0.10  &0.03      \\
Po       &  84  & 221 &   1697.62      &1696.21 &0.34 & 0.12  &1.41      \\
Po       &  84  & 222 &   1702.98      &1702.97 &1.54 & 0.11  &0.01      \\
At       &  85  & 220 &   1694.15      &1694.33 &0.45 & 0.12  &-0.18      \\
At       &  85  & 221 &   1699.82      &1701.19 &1.47 & 0.12  &-1.37      \\
At       &  85  & 222 &   1703.72      &1703.84 &0.55 & 0.13  &-0.12      \\
At       &  85  & 223 &   1709.31      &1710.26 &1.21 & 0.13  &-0.95      \\
At       &  85  & 224 &   1713.10      &1712.53 &0.17 & 0.14  &0.57      \\
Rn       &  86  & 223 &   1711.56      &1711.44 &0.67 & 0.15  &0.12      \\
Rn       &  86  & 224 &   1717.56      &1718.35 &1.59 & 0.15  &-0.79      \\
Rn       &  86  & 225 &   1721.56      &1720.80 &0.41 & 0.16  &0.76      \\
Rn       &  86  & 226 &   1727.40      &1728.04 &1.71 & 0.16  &-0.64      \\
Rn       &  86  & 227 &   1731.32      &1730.65 &0.81 & 0.18  &0.67      \\
Rn       &  86  & 228 &   1737.06      &1737.41 &1.80 & 0.17  &-0.35      \\
Fr       &  87  & 224 &   1717.41      &1716.91 &0.42 & 0.14  &0.50      \\
Fr       &  87  & 225 &   1723.46      &1724.75 &1.79 & 0.14  &-1.29      \\
Fr       &  87  & 226 &   1727.81      &1727.27 &0.64 & 0.15  &0.54      \\
Fr       &  87  & 227 &   1733.74      &1734.87 &1.90 & 0.16  &-1.13      \\
Fr       &  87  & 228 &   1738.13      &1737.28 &0.64 & 0.17  &0.85      \\
Fr       &  87  & 229 &   1743.89      &1744.20 &1.93 & 0.20  &-0.31      \\
Fr       &  87  & 230 &   1748.12      &1747.22 &0.92 & 0.18  &0.90      \\
Fr       &  87  & 231 &   1753.64      &1753.79 &2.05 & 0.21  &-0.15      \\
Ra       &  88  & 231 &   1756.69      &1756.10 &1.81 & 0.24  &0.59      \\
Ra       &  88  & 232 &   1762.48      &1762.41 &2.27 & 0.24  &0.07      \\
Ra       &  88  & 233 &   1766.72      &1765.72 &1.69 & 0.26  &1.00      \\
Ra       &  88  & 234 &   1772.23      &1772.06 &2.26 & 0.25  &0.16      \\
Ac       &  89  & 229 &   1747.27      &1747.36 &1.88 & 0.23  &-0.09      \\
Ac       &  89  & 230 &   1752.19      &1751.21 &1.12 & 0.24  &0.98      \\
Ac       &  89  & 231 &   1758.34      &1758.28 &1.90 & 0.24  &0.06      \\
Ac       &  89  & 232 &   1763.02      &1762.50 &0.66 & 0.26  &0.52      \\
Ac       &  89  & 233 &   1768.94      &1768.69 &1.90 & 0.25  &0.25      \\
Ac       &  89  & 234 &   1773.47      &1772.41 &1.52 & 0.27  &1.06      \\
Ac       &  89  & 235 &   1779.03      &1778.66 &1.88 & 0.26  &0.37      \\
Ac       &  89  & 236 &   1783.24      &1782.32 &1.52 & 0.27  &0.92      \\
Th       &  90  & 235 &   1781.57      &1780.76 &1.80 & 0.28  &0.81      \\
Th       &  90  & 236 &   1787.40      &1787.27 &2.27 & 0.27  &0.13      \\
Th       &  90  & 237 &   1791.77      &1791.08 &1.84 & 0.28  &0.69      \\
Pa       &  91  & 235 &   1782.49      &1782.02 &1.40 & 0.27  &0.47      \\
Pa       &  91  & 236 &   1787.52      &1786.25 &1.12 & 0.28  &1.27      \\
Pa       &  91  & 237 &   1793.40      &1792.37 &1.03 & 0.27  &1.03      \\
Pa       &  91  & 238 &   1798.10      &1796.41 &0.73 & 0.28  &1.69      \\
\hline
\noalign{\vspace{5pt}}
$\Delta$ &      &  &   &   &  & & 0.859    \\
\end{longtable}

\begin{figure*}[!htbp]
\includegraphics[width=12cm]{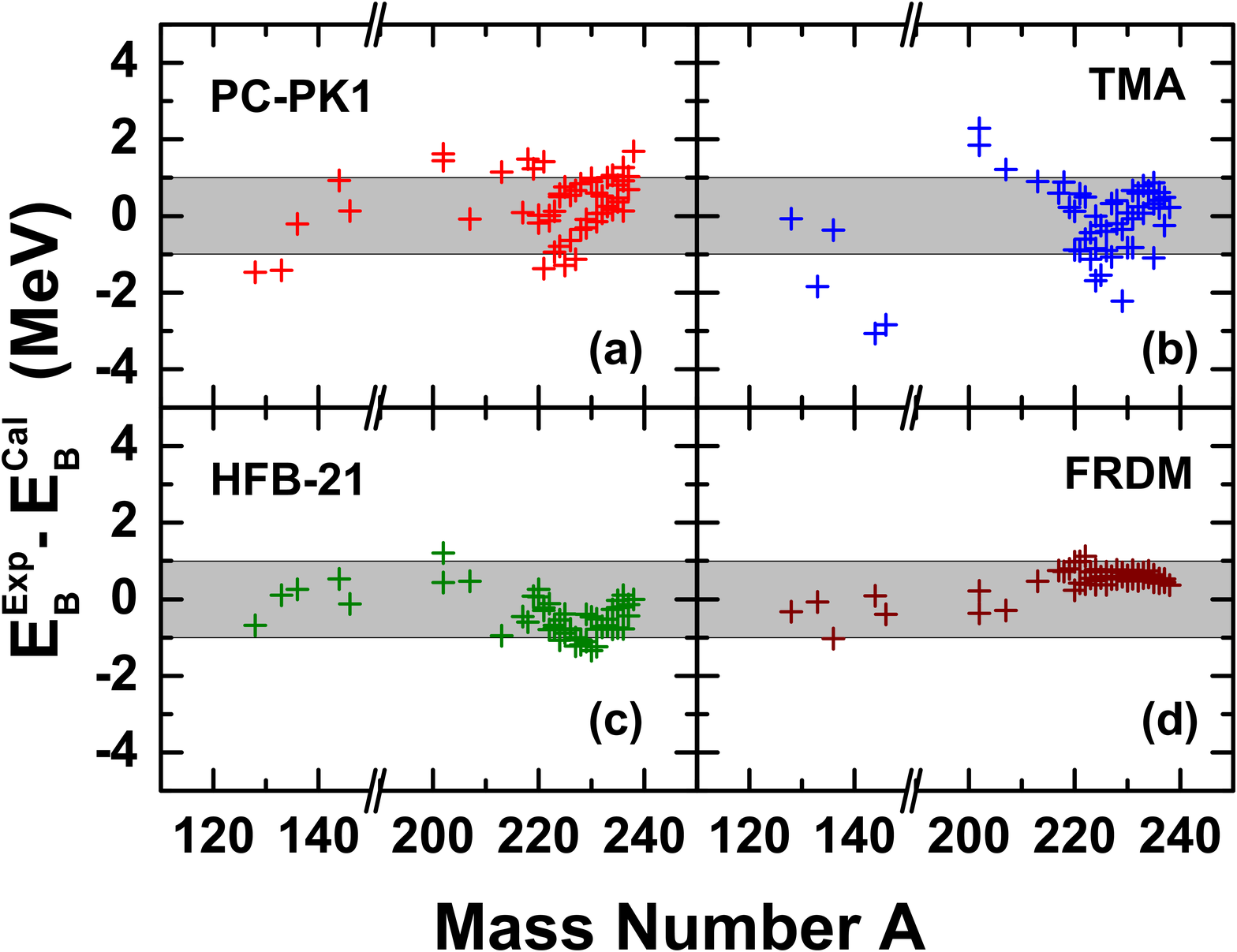}
\caption{(Color online) Differences between the experimental data~\cite{Chen2012Nucl.Phys.A71} and the calculated binding energies by the covariant density functional theory (CDFT) with the effective interaction PC-PK1~\cite{Zhao2010Phys.Rev.C54319} and TMA~\cite{Geng2005Prog.Theor.Phys.785} as well as those from the mass models HFB-21~\cite{Goriely2010Phys.Rev.C35804} and FRDM~\cite{Moller1995At.DataNucl.DataTables185}.}
\label{fig1}
\end{figure*}

In order to present the calculated results more plainly, in Fig.~\ref{fig1}, we show the differences between the experimental data~\cite{Chen2012Nucl.Phys.A71} and the calculated binding energies with the PC-PK1~\cite{Zhao2010Phys.Rev.C54319} in comparison with those obtained from the CDFT calculations with TMA~\cite{Geng2005Prog.Theor.Phys.785} as well as the mass models HFB-21~\cite{Goriely2010Phys.Rev.C35804} and FRDM~\cite{Moller1995At.DataNucl.DataTables185}.

It is found that the deviations given by both PC-PK1 and TMA are less than 1 MeV for most nuclei. Both cases underestimate the binding energies for nuclei near $N=126$ and overestimate them for nuclei near $N=82$. Moreover, the deviations for nuclei $^{144}$La and $^{146}$Ce in TMA are even as large as 3 MeV. This indicates that PC-PK1 is more robust in the predictions of nuclear masses, especially for the neutron-rich nuclei. The up-to-date nonrelativistic mass model HFB-21~\cite{Goriely2010Phys.Rev.C35804} could also reproduce the data  quite well. However, it should be noted that there are more than 20 parameters in HFB-21 including the phenomenological corrections, and they are determined by constraining the nuclear-matter parameters and optimizing the fit to the full data set of the 2149 measured masses with $N,Z \ge 8$ (both spherical and deformed)~\cite{Audi2003Nucl.Phys.A337}, including some of the present data (improved masses in the work~\cite{Chen2012Nucl.Phys.A71}), while the functional PC-PK1 contains only 11 parameters, which are determined by fitting to the binding energies of only 60 selected spherical nuclei and charge radii for 17 ones~\cite{Zhao2010Phys.Rev.C54319}. It should be mentioned that the macroscopic-microscopic FRDM also gives good agreement with the data.

Focusing on the heavy mass region in Fig.~\ref{fig1}, especially for the nuclei with $A>210$, one can clearly see some systematic regularities for the mass deviations. Both the PC-PK1 and TMA results show linear-like increasing tendency with the mass number, and the HFB-21 results exhibit the opposite tendency. In addition, the results given by PC-PK1 could be roughly classified into two branches. For FRDM, all the deviations for the heavy mass region are concentrated around 0.6 MeV.

\begin{figure*}[!htbp]
\includegraphics[width=12cm]{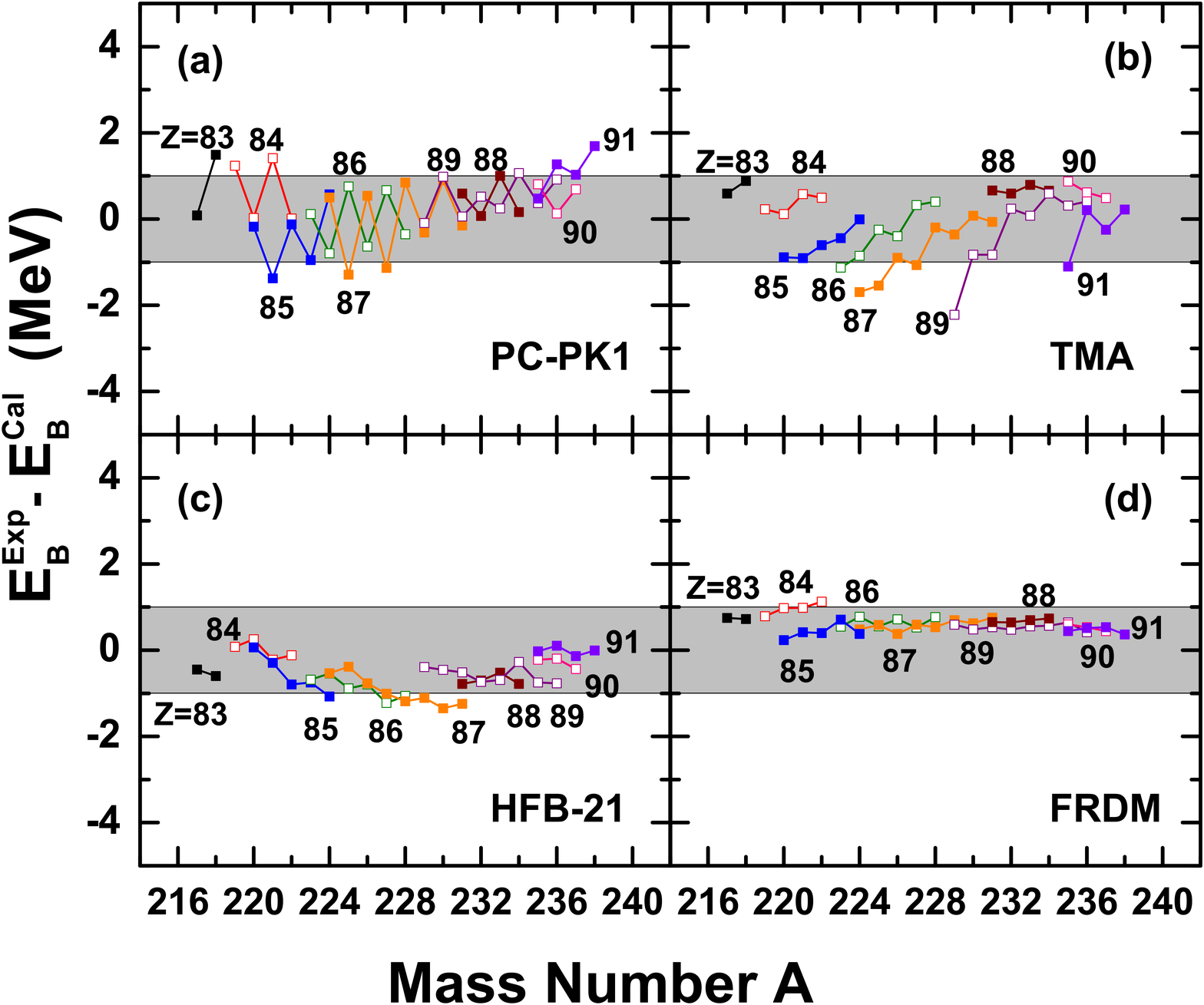}
\caption{(Color online) Same as Fig.~\ref{fig1} but for the isotopic chains from $Z=83$ to $Z=91$.}
\label{fig2}
\end{figure*}

In order to investigate this in more detail, in Fig.~\ref{fig2}, differences between the data and the calculated binding energies for nine isotopic chains from $Z=83$ to $Z=91$ are shown. It can be seen that the deviations for FRDM remain constant with mass number. For both PC-PK1 and TMA, the binding energy differences of most isotopic chains grow steadily with the neutron number, which indicates that the isotopes in most isotopic chains become more underbound with the increase of the neutron number. In contrast, for the mass model HFB-21, the isotopes in most isotopic chains become more overbound with the increase of the neutron number. Such difference may result from the different recipes employed in the calculations for treating the pairing correlations. In the CDFT calculations with PC-PK1 and TMA, the pairing correlation is treated with the simple BCS method, while in the mass model HFB-21, it is treated with the sophisticated Bogoliubov transformation. For the neutron-rich nuclei, the latter could describe the pairing correlation with the continuum properly and therefore include more correlations which makes the nucleus more bound. It was found in Ref.~\cite{Samyn2002Nucl.Phys.A142} that there is a general tendency for overbinding with the HFB calculations in comparison with the HF + BCS case, and the corresponding mass deviations could be up to 1.5 MeV for open-shell nuclei. Therefore, future efforts should be devoted to the sophisticated and heavy Bogoliubov calculations with continuum for the neutron-rich heavy nuclei as shown in Refs.~\cite{Meng1996Phys.Rev.Lett.3963,Meng1998Nucl.Phys.A3} for the spherical case and Refs.~\cite{Zhou2010Phys.Rev.C011301R,Li2012Phys.Rev.C24312,Chen2012Phys.Rev.C67301} for the deformed case.

\begin{figure}[!htbp]
\includegraphics[width=8.5cm]{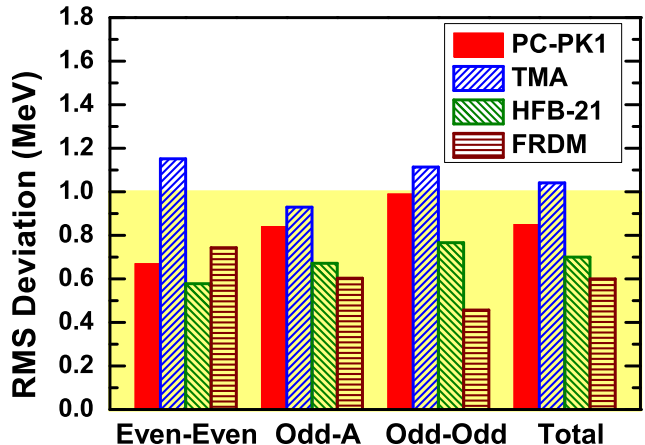}
\caption{(Color online) The rms deviations of the calculated binding energies by the CDFT with the effective interaction PC-PK1 and TMA as well as those from the mass models HFB-21 and FRDM for the even-even, odd-$A$, odd-odd, and the whole set of nuclei.}
\label{fig3}
\end{figure}

Comparing the results given by PC-PK1 and TMA in Fig.~\ref{fig2}, one could see that there are noticeable odd-even staggering in the results given by PC-PK1, which eventually leads to the two linear-like branches of the deviations.
This staggering phenomenon influences the rms value to a great extent.
In Fig.~\ref{fig3}, the rms deviations of the calculated binding energies by the CDFT with PC-PK1 and TMA as well as those from the mass models HFB-21 and FRDM are shown for the even-even, odd-$A$, odd-odd, and the whole set of nuclei, respectively. For the whole set of nuclei, the rms deviation given by PC-PK1 is about 0.8 MeV. This is larger than the rms value 0.7 MeV given by HFB-21 but smaller than the value of 1.0 MeV given by TMA. Specifically, for the 12 even-even nuclei, PC-PK1 achieves very good agreement with the data and the rms deviation is about 0.6 MeV. This is at the same good level as the HFB-21 mass model and much better than that given by TMA. However, for the 25 odd-$A$ nuclei and 16 odd-odd nuclei, the rms values given by PC-PK1 become larger but still within 1 MeV. In contrast to the microscopic frameworks, the mass model FRDM gives very good agreement for the odd-odd nuclei with a rms deviation less than 0.5 MeV. This lowers the corresponding rms deviation for the whole set of nuclei to 0.6 MeV.

\begin{figure}[!htbp]
\includegraphics[width=8.5cm]{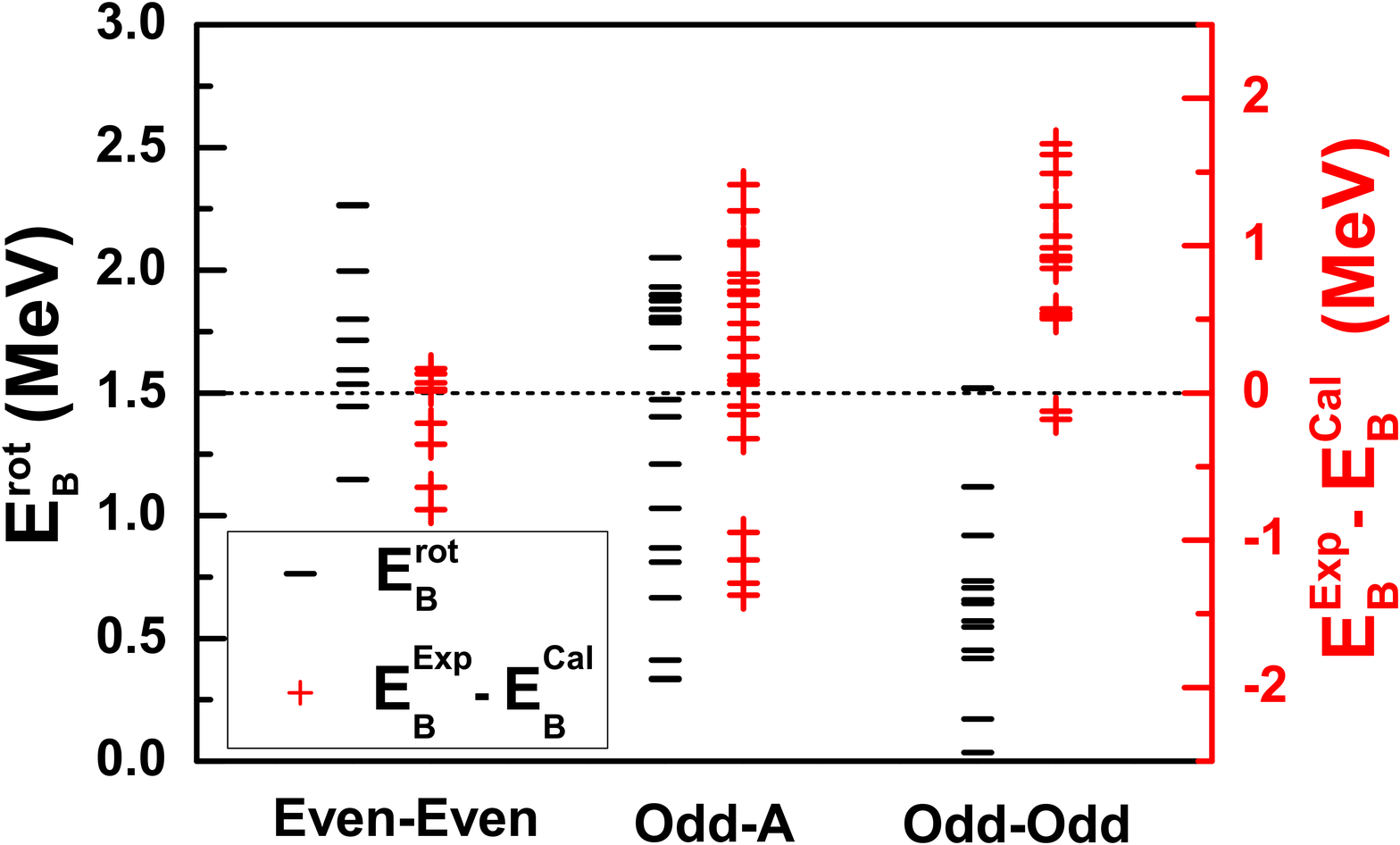}
\caption{(Color online) Correlations between the deviations of the calculated binding energies with PC-PK1 from the data and the calculated rotational correction energies for even-even, odd-$A$, and odd-odd nuclei. Note that the order in different data sets is independent.}
\label{fig4}
\end{figure}

To explore the reason for this enhancement of the rms values for odd particle systems, in Fig.~\ref{fig4}, correlations between the deviations of the calculated binding energies with PC-PK1 from the data and the calculated rotational correction energies are shown for even-even, odd-$A$, and odd-odd nuclei respectively. It is clear that the rotational correction energy plays an important role in the determination of the nuclear masses with PC-PK1. In particular, for the odd-odd nuclei, the rotation correction energies are systemically smaller than those of the even-even nuclei even if they have similar deformations. Apparently, this leads to the systematical underestimation of the binding energies for odd-odd nuclei.

This may result from two reasons. On one hand, for the odd particle system, the odd nucleon breaks the time-reversal symmetry and the time-odd components of the vector potential do not vanish. Although it is generally believed that the contributions due to the time-reversal symmetry breaking contribute little to the bulk properties such as binding energies, they may significantly influence features like magnetic moment~\cite{Yao2006Phys.Rev.C24307} or moment of inertia~\cite{Koepf1990Nucl.Phys.A279}, which depend on a few valence nucleons. Here, the calculation of rotation correction energy  with Eq.~(\ref{Erot}) involves with the moment of inertia, which is calculated without the time-odd components. Therefore, the neglect of the time-odd components may affect the rotation correction energy.

On the other hand, for odd particle systems, the unpaired particle will block its occupied level in the BCS calculations; that is the Pauli principle prevents this level from the scattering process of nucleon pairs by the pairing correlations. In principle, the ground-state for odd particle systems should be the state with the lowest energy determined by calculations blocking each possible level near the Fermi surface.
However, this is incredibly time-consuming.

Here the blocking method adopted in the BCS method is the following: For each step of the self-consistent calculation, the level to be blocked is determined by the filling approximation from the single-particle spectra obtained. Of course, it may occur that the state chosen by this procedure is not the true ground state. As pointed out in Ref.~\cite{Geng2005}, such wrong selection of the blocking state generally leads to a difference less than 0.2 MeV for the binding energy. However, this would remarkably influence the moment of inertia and thus the rotation correction energy. Therefore, further improvement in the evaluation of the rotation correction energy, in particular the moments of inertia, would be needed in the framework of CDFT.

In summary, the CDFT with the point-coupling interaction PC-PK1 is applied to investigate the new experimental masses  of 53 heavy neutron-rich isotopes in the element range of 50 to 91. The experimental data are from a mass measurement performed with the storage ring mass spectrometry at GSI~\cite{Chen2012Nucl.Phys.A71}. The functional PC-PK1 contains only 11 parameters, which are determined by fitting to the binding energies of 60 spherical nuclei and charge radii for 17 ones. It is found that the CDFT with PC-PK1 can reproduce the experimental data quite well and the corresponding rms deviation is 0.859 MeV.
For the 25 odd-$A$ nuclei and 16 odd-odd nuclei, the rms values given by PC-PK1 are a little large but still within 1 MeV, which may be improved in the future by treating properly the time-odd component and the moment of inertia. An excellent predictive power with an accuracy of about 600 keV has been achieved for even-even nuclei with the PC-PK1.

\begin{acknowledgments}
This work was supported in part by the Major State 973 Program of China (Grant No. 2013CB834400), the Natural Science Foundation of China (Grants No. 10975007, No. 10975008, No. 11175002, No. 11105005, No. 11105010, No. 11035007, No. 11128510, No. 11235002), and the Research Fund for the Doctoral Program of Higher Education under Grant No. 20110001110087 and the China Postdoctoral Science Foundation Grant No. 2012M520101. B.S. is also partially supported by NCET-09-0031.
\end{acknowledgments}


\begin{thebibliography}{62}
\expandafter\ifx\csname natexlab\endcsname\relax\def\natexlab#1{#1}\fi
\expandafter\ifx\csname bibnamefont\endcsname\relax
  \def\bibnamefont#1{#1}\fi
\expandafter\ifx\csname bibfnamefont\endcsname\relax
  \def\bibfnamefont#1{#1}\fi
\expandafter\ifx\csname citenamefont\endcsname\relax
  \def\citenamefont#1{#1}\fi
\expandafter\ifx\csname url\endcsname\relax
  \def\url#1{\texttt{#1}}\fi
\expandafter\ifx\csname urlprefix\endcsname\relax\def\urlprefix{URL }\fi
\providecommand{\bibinfo}[2]{#2}
\providecommand{\eprint}[2][]{\url{#2}}

\bibitem[{\citenamefont{Lunney et~al.}(2003)\citenamefont{Lunney, Pearson, and
  Thibault}}]{Lunney2003Rev.Mod.Phys.1021}
\bibinfo{author}{\bibfnamefont{D.}~\bibnamefont{Lunney}},
  \bibinfo{author}{\bibfnamefont{J.~M.} \bibnamefont{Pearson}},
  \bibnamefont{and} \bibinfo{author}{\bibfnamefont{C.}~\bibnamefont{Thibault}},
  \bibinfo{journal}{Rev. Mod. Phys.} \textbf{\bibinfo{volume}{75}},
  \bibinfo{pages}{1021} (\bibinfo{year}{2003}).

\bibitem[{\citenamefont{von Weizsacker}(1935)}]{Weizsacker1935Z.Physik431}
\bibinfo{author}{\bibfnamefont{C.~F.} \bibnamefont{von Weizsacker}},
  \bibinfo{journal}{Z. Phys.} \textbf{\bibinfo{volume}{96}},
  \bibinfo{pages}{431} (\bibinfo{year}{1935}).

\bibitem[{\citenamefont{M\"oller et~al.}(1995)\citenamefont{M\"oller, Nix,
  Myers, and Swiatecki}}]{Moller1995At.DataNucl.DataTables185}
\bibinfo{author}{\bibfnamefont{P.}~\bibnamefont{M\"oller}},
  \bibinfo{author}{\bibfnamefont{J.~R.} \bibnamefont{Nix}},
  \bibinfo{author}{\bibfnamefont{W.~D.} \bibnamefont{Myers}}, \bibnamefont{and}
  \bibinfo{author}{\bibfnamefont{W.~J.} \bibnamefont{Swiatecki}},
  \bibinfo{journal}{At. Data Nucl. Data Tables} \textbf{\bibinfo{volume}{59}},
  \bibinfo{pages}{185 } (\bibinfo{year}{1995}).

\bibitem[{\citenamefont{Pearson et~al.}(1996)\citenamefont{Pearson, Nayak, and
  Goriely}}]{Pearson1996Phys.Lett.B455}
\bibinfo{author}{\bibfnamefont{J.~M.} \bibnamefont{Pearson}},
  \bibinfo{author}{\bibfnamefont{R.~C.} \bibnamefont{Nayak}}, \bibnamefont{and}
  \bibinfo{author}{\bibfnamefont{S.}~\bibnamefont{Goriely}},
  \bibinfo{journal}{Phys. Lett. B} \textbf{\bibinfo{volume}{387}},
  \bibinfo{pages}{455} (\bibinfo{year}{1996}).

\bibitem[{\citenamefont{Wang et~al.}(2010)\citenamefont{Wang, Liang, Liu, and
  Wu}}]{Wang2010Phys.Rev.C44304}
\bibinfo{author}{\bibfnamefont{N.}~\bibnamefont{Wang}},
  \bibinfo{author}{\bibfnamefont{Z.}~\bibnamefont{Liang}},
  \bibinfo{author}{\bibfnamefont{M.}~\bibnamefont{Liu}}, \bibnamefont{and}
  \bibinfo{author}{\bibfnamefont{X.}~\bibnamefont{Wu}}, \bibinfo{journal}{Phys.
  Rev. C} \textbf{\bibinfo{volume}{82}}, \bibinfo{pages}{044304}
  (\bibinfo{year}{2010}).

\bibitem[{\citenamefont{Ormand}(1997)}]{Ormand1997Phys.Rev.C2407}
\bibinfo{author}{\bibfnamefont{W.~E.} \bibnamefont{Ormand}},
  \bibinfo{journal}{Phys. Rev. C} \textbf{\bibinfo{volume}{55}},
  \bibinfo{pages}{2407} (\bibinfo{year}{1997}).

\bibitem[{\citenamefont{Barea et~al.}(2008)\citenamefont{Barea, Frank, Hirsch,
  Isacker, Pittel, and Vel\'azquez}}]{Barea2008Phys.Rev.C41304}
\bibinfo{author}{\bibfnamefont{J.}~\bibnamefont{Barea}},
  \bibinfo{author}{\bibfnamefont{A.}~\bibnamefont{Frank}},
  \bibinfo{author}{\bibfnamefont{J.~G.} \bibnamefont{Hirsch}},
  \bibinfo{author}{\bibfnamefont{P.~Van} \bibnamefont{Isacker}},
  \bibinfo{author}{\bibfnamefont{S.}~\bibnamefont{Pittel}}, \bibnamefont{and}
  \bibinfo{author}{\bibfnamefont{V.}~\bibnamefont{Vel\'azquez}},
  \bibinfo{journal}{Phys. Rev. C} \textbf{\bibinfo{volume}{77}},
  \bibinfo{pages}{041304} (\bibinfo{year}{2008}).

\bibitem[{\citenamefont{Fu et~al.}(2011)\citenamefont{Fu, Lei, Jiang, Zhao,
  Sun, and Arima}}]{Fu2011Phys.Rev.C34311}
\bibinfo{author}{\bibfnamefont{G.~J.} \bibnamefont{Fu}},
  \bibinfo{author}{\bibfnamefont{Y.}~\bibnamefont{Lei}},
  \bibinfo{author}{\bibfnamefont{H.}~\bibnamefont{Jiang}},
  \bibinfo{author}{\bibfnamefont{Y.~M.} \bibnamefont{Zhao}},
  \bibinfo{author}{\bibfnamefont{B.}~\bibnamefont{Sun}}, \bibnamefont{and}
  \bibinfo{author}{\bibfnamefont{A.}~\bibnamefont{Arima}},
  \bibinfo{journal}{Phys. Rev. C} \textbf{\bibinfo{volume}{84}},
  \bibinfo{pages}{034311} (\bibinfo{year}{2011}).

\bibitem[{\citenamefont{Goriely
  et~al.}(2009{\natexlab{a}})\citenamefont{Goriely, Chamel, and
  Pearson}}]{Goriely2009Phys.Rev.Lett.152503}
\bibinfo{author}{\bibfnamefont{S.}~\bibnamefont{Goriely}},
  \bibinfo{author}{\bibfnamefont{N.}~\bibnamefont{Chamel}}, \bibnamefont{and}
  \bibinfo{author}{\bibfnamefont{J.~M.} \bibnamefont{Pearson}},
  \bibinfo{journal}{Phys. Rev. Lett.} \textbf{\bibinfo{volume}{102}},
  \bibinfo{eid}{152503} (\bibinfo{year}{2009}{\natexlab{a}}).

\bibitem[{\citenamefont{Goriely
  et~al.}(2009{\natexlab{b}})\citenamefont{Goriely, Hilaire, Girod, and
  P\'eru}}]{Goriely2009Phys.Rev.Lett.242501}
\bibinfo{author}{\bibfnamefont{S.}~\bibnamefont{Goriely}},
  \bibinfo{author}{\bibfnamefont{S.}~\bibnamefont{Hilaire}},
  \bibinfo{author}{\bibfnamefont{M.}~\bibnamefont{Girod}}, \bibnamefont{and}
  \bibinfo{author}{\bibfnamefont{S.}~\bibnamefont{P\'eru}},
  \bibinfo{journal}{Phys. Rev. Lett.} \textbf{\bibinfo{volume}{102}},
  \bibinfo{pages}{242501} (\bibinfo{year}{2009}{\natexlab{b}}).

\bibitem[{\citenamefont{Goriely et~al.}(2010)\citenamefont{Goriely, Chamel, and
  Pearson}}]{Goriely2010Phys.Rev.C35804}
\bibinfo{author}{\bibfnamefont{S.}~\bibnamefont{Goriely}},
  \bibinfo{author}{\bibfnamefont{N.}~\bibnamefont{Chamel}}, \bibnamefont{and}
  \bibinfo{author}{\bibfnamefont{J.~M.} \bibnamefont{Pearson}},
  \bibinfo{journal}{Phys. Rev. C} \textbf{\bibinfo{volume}{82}},
  \bibinfo{pages}{035804} (\bibinfo{year}{2010}).

\bibitem[{\citenamefont{Ring}(1996)}]{Ring1996Prog.Part.Nucl.Phys.193}
\bibinfo{author}{\bibfnamefont{P.}~\bibnamefont{Ring}}, \bibinfo{journal}{Prog.
  Part. Nucl. Phys.} \textbf{\bibinfo{volume}{37}}, \bibinfo{pages}{193 }
  (\bibinfo{year}{1996}).

\bibitem[{\citenamefont{Meng et~al.}(2006)\citenamefont{Meng, Toki, Zhou,
  Zhang, Long, and Geng}}]{Meng2006Prog.Part.Nucl.Phys.470}
\bibinfo{author}{\bibfnamefont{J.}~\bibnamefont{Meng}},
  \bibinfo{author}{\bibfnamefont{H.}~\bibnamefont{Toki}},
  \bibinfo{author}{\bibfnamefont{S.}~\bibnamefont{Zhou}},
  \bibinfo{author}{\bibfnamefont{S.}~\bibnamefont{Zhang}},
  \bibinfo{author}{\bibfnamefont{W.}~\bibnamefont{Long}}, \bibnamefont{and}
  \bibinfo{author}{\bibfnamefont{L.}~\bibnamefont{Geng}},
  \bibinfo{journal}{Prog. Part. Nucl. Phys.} \textbf{\bibinfo{volume}{57}},
  \bibinfo{pages}{470} (\bibinfo{year}{2006}).

\bibitem[{\citenamefont{Vretenar et~al.}(2005)\citenamefont{Vretenar,
  Afanasjev, and Ring}}]{Vretenar2005Phys.Rep.101}
\bibinfo{author}{\bibfnamefont{D.}~\bibnamefont{Vretenar}},
  \bibinfo{author}{\bibfnamefont{A. V.}~\bibnamefont{Afanasjev},
  \bibfnamefont{G. A.~Lalazissis}}, \bibnamefont{and}
  \bibinfo{author}{\bibfnamefont{P.}~\bibnamefont{Ring}},
  \bibinfo{journal}{Phys. Rep.} \textbf{\bibinfo{volume}{409}},
  \bibinfo{pages}{101} (\bibinfo{year}{2005}).

\bibitem[{\citenamefont{Nik\v{s}i\'{c}
  et~al.}(2011)\citenamefont{Nik\v{s}i\'{c}, Vretenar, and
  Ring}}]{Niksic2011Prog.Part.Nucl.Phys.519}
\bibinfo{author}{\bibfnamefont{T.}~\bibnamefont{Nik\v{s}i\'{c}}},
  \bibinfo{author}{\bibfnamefont{D.}~\bibnamefont{Vretenar}}, \bibnamefont{and}
  \bibinfo{author}{\bibfnamefont{P.}~\bibnamefont{Ring}},
  \bibinfo{journal}{Prog. Part. Nucl. Phys.} \textbf{\bibinfo{volume}{66}},
  \bibinfo{pages}{519 } (\bibinfo{year}{2011}).

\bibitem[{\citenamefont{Ginocchio}(2005)}]{Ginocchio2005Phys.Rep.165}
\bibinfo{author}{\bibfnamefont{J.~N.} \bibnamefont{Ginocchio}},
  \bibinfo{journal}{Phys. Rep.} \textbf{\bibinfo{volume}{414}},
  \bibinfo{pages}{165} (\bibinfo{year}{2005}).

\bibitem[{\citenamefont{Meng et~al.}(1999)\citenamefont{Meng, Sugawara-Tanabe,
  Yamaji, and Arima}}]{Meng1999Phys.Rev.C154}
\bibinfo{author}{\bibfnamefont{J.}~\bibnamefont{Meng}},
  \bibinfo{author}{\bibfnamefont{K.}~\bibnamefont{Sugawara-Tanabe}},
  \bibinfo{author}{\bibfnamefont{S.}~\bibnamefont{Yamaji}}, \bibnamefont{and}
  \bibinfo{author}{\bibfnamefont{A.}~\bibnamefont{Arima}},
  \bibinfo{journal}{Phys. Rev. C} \textbf{\bibinfo{volume}{59}},
  \bibinfo{pages}{154} (\bibinfo{year}{1999}).

\bibitem[{\citenamefont{Meng et~al.}(1998)\citenamefont{Meng, Sugawara-Tanabe,
  Yamaji, Ring, and Arima}}]{Meng1998Phys.Rev.C628}
\bibinfo{author}{\bibfnamefont{J.}~\bibnamefont{Meng}},
  \bibinfo{author}{\bibfnamefont{K.}~\bibnamefont{Sugawara-Tanabe}},
  \bibinfo{author}{\bibfnamefont{S.}~\bibnamefont{Yamaji}},
  \bibinfo{author}{\bibfnamefont{P.}~\bibnamefont{Ring}}, \bibnamefont{and}
  \bibinfo{author}{\bibfnamefont{A.}~\bibnamefont{Arima}},
  \bibinfo{journal}{Phys. Rev. C} \textbf{\bibinfo{volume}{58}},
  \bibinfo{pages}{R628} (\bibinfo{year}{1998}).

\bibitem[{\citenamefont{Liang et~al.}(2011)\citenamefont{Liang, Zhao, Zhang,
  Meng, and Giai}}]{Liang2011Phys.Rev.C41301}
\bibinfo{author}{\bibfnamefont{H.}~\bibnamefont{Liang}},
  \bibinfo{author}{\bibfnamefont{P.}~\bibnamefont{Zhao}},
  \bibinfo{author}{\bibfnamefont{Y.}~\bibnamefont{Zhang}},
  \bibinfo{author}{\bibfnamefont{J.}~\bibnamefont{Meng}}, \bibnamefont{and}
  \bibinfo{author}{\bibfnamefont{N.~V.} \bibnamefont{Giai}},
  \bibinfo{journal}{Phys. Rev. C} \textbf{\bibinfo{volume}{83}},
  \bibinfo{pages}{041301} (\bibinfo{year}{2011}).

\bibitem[{\citenamefont{Guo}(2012)}]{Guo2012Phys.Rev.C21302}
\bibinfo{author}{\bibfnamefont{J.-Y.} \bibnamefont{Guo}},
  \bibinfo{journal}{Phys. Rev. C} \textbf{\bibinfo{volume}{85}},
  \bibinfo{pages}{021302} (\bibinfo{year}{2012}).

\bibitem[{\citenamefont{Lu et~al.}(2012{\natexlab{a}})\citenamefont{Lu, Zhao,
  and Zhou}}]{Lu2012Phys.Rev.Lett.72501}
\bibinfo{author}{\bibfnamefont{B.-N.} \bibnamefont{Lu}},
  \bibinfo{author}{\bibfnamefont{E.-G.} \bibnamefont{Zhao}}, \bibnamefont{and}
  \bibinfo{author}{\bibfnamefont{S.-G.} \bibnamefont{Zhou}},
  \bibinfo{journal}{Phys. Rev. Lett.} \textbf{\bibinfo{volume}{109}},
  \bibinfo{pages}{072501} (\bibinfo{year}{2012}{\natexlab{a}}).

\bibitem[{\citenamefont{Zhou et~al.}(2003)\citenamefont{Zhou, Meng, and
  Ring}}]{Zhou2003Phys.Rev.Lett.262501}
\bibinfo{author}{\bibfnamefont{S.-G.} \bibnamefont{Zhou}},
  \bibinfo{author}{\bibfnamefont{J.}~\bibnamefont{Meng}}, \bibnamefont{and}
  \bibinfo{author}{\bibfnamefont{P.}~\bibnamefont{Ring}},
  \bibinfo{journal}{Phys. Rev. Lett.} \textbf{\bibinfo{volume}{91}},
  \bibinfo{pages}{262501} (\bibinfo{year}{2003}).

\bibitem[{\citenamefont{Koepf and Ring}(1989)}]{Koepf1989Nucl.Phys.A61}
\bibinfo{author}{\bibfnamefont{W.}~\bibnamefont{Koepf}} \bibnamefont{and}
  \bibinfo{author}{\bibfnamefont{P.}~\bibnamefont{Ring}},
  \bibinfo{journal}{Nucl. Phys. A} \textbf{\bibinfo{volume}{493}},
  \bibinfo{pages}{61 } (\bibinfo{year}{1989}).

\bibitem[{\citenamefont{Yao et~al.}(2006)\citenamefont{Yao, Chen, and
  Meng}}]{Yao2006Phys.Rev.C24307}
\bibinfo{author}{\bibfnamefont{J.~M.} \bibnamefont{Yao}},
  \bibinfo{author}{\bibfnamefont{H.}~\bibnamefont{Chen}}, \bibnamefont{and}
  \bibinfo{author}{\bibfnamefont{J.}~\bibnamefont{Meng}},
  \bibinfo{journal}{Phys. Rev. C} \textbf{\bibinfo{volume}{74}},
  \bibinfo{eid}{024307} (\bibinfo{year}{2006}).

\bibitem[{\citenamefont{Arima}(2011)}]{Arima2011SciChinaSerG-PhysMechAstron188}
\bibinfo{author}{\bibfnamefont{A.}~\bibnamefont{Arima}}, \bibinfo{journal}{Sci. China: Phys. Mech. Astron.} \textbf{\bibinfo{volume}{54}},
  \bibinfo{pages}{188} (\bibinfo{year}{2011}).

\bibitem[{\citenamefont{Li et~al.}(2011{\natexlab{a}})\citenamefont{Li, Meng,
  Ring, Yao, and Arima}}]{Li2011Sci.ChinaPhys.Mech.Astron.204}
\bibinfo{author}{\bibfnamefont{J.}~\bibnamefont{Li}},
  \bibinfo{author}{\bibfnamefont{J.}~\bibnamefont{Meng}},
  \bibinfo{author}{\bibfnamefont{P.}~\bibnamefont{Ring}},
  \bibinfo{author}{\bibfnamefont{J.~M.} \bibnamefont{Yao}}, \bibnamefont{and}
  \bibinfo{author}{\bibfnamefont{A.}~\bibnamefont{Arima}},
  \bibinfo{journal}{Sci. China: Phys. Mech. Astron.}
  \textbf{\bibinfo{volume}{54}}, \bibinfo{pages}{204}
  (\bibinfo{year}{2011}{\natexlab{a}}).

\bibitem[{\citenamefont{Li et~al.}(2011{\natexlab{b}})\citenamefont{Li, Yao,
  Meng, and Arima}}]{Li2011Prog.Theor.Phys.1185}
\bibinfo{author}{\bibfnamefont{J.}~\bibnamefont{Li}},
  \bibinfo{author}{\bibfnamefont{J.~M.} \bibnamefont{Yao}},
  \bibinfo{author}{\bibfnamefont{J.}~\bibnamefont{Meng}}, \bibnamefont{and}
  \bibinfo{author}{\bibfnamefont{A.}~\bibnamefont{Arima}},
  \bibinfo{journal}{Prog. Theor. Phys.} \textbf{\bibinfo{volume}{125}},
  \bibinfo{pages}{1185} (\bibinfo{year}{2011}{\natexlab{b}}).

\bibitem[{\citenamefont{Wei et~al.}(2012)\citenamefont{Wei, Li, and
  Meng}}]{Wei2012Prog.Theor.Phys.400}
\bibinfo{author}{\bibfnamefont{J.}~\bibnamefont{Wei}},
  \bibinfo{author}{\bibfnamefont{J.}~\bibnamefont{Li}}, \bibnamefont{and}
  \bibinfo{author}{\bibfnamefont{J.}~\bibnamefont{Meng}},
  \bibinfo{journal}{Prog. Theor. Phys.} \textbf{\bibinfo{volume}{S196}}, \bibinfo{pages}{400} (\bibinfo{year}{2012}).

\bibitem[{\citenamefont{Afanasjev et~al.}(2000)\citenamefont{Afanasjev, Ring,
  and K\"onig}}]{Afanasjev2000Nucl.Phys.A196}
\bibinfo{author}{\bibfnamefont{A.~V.} \bibnamefont{Afanasjev}},
  \bibinfo{author}{\bibfnamefont{P.}~\bibnamefont{Ring}}, \bibnamefont{and}
  \bibinfo{author}{\bibfnamefont{J.}~\bibnamefont{K\"onig}},
  \bibinfo{journal}{Nucl. Phys. A} \textbf{\bibinfo{volume}{676}},
  \bibinfo{pages}{196 } (\bibinfo{year}{2000}).

\bibitem[{\citenamefont{Zhao et~al.}(2012)\citenamefont{Zhao, Peng, Liang,
  Ring, and Meng}}]{Zhao2012Phys.Rev.C54310}
\bibinfo{author}{\bibfnamefont{P.~W.} \bibnamefont{Zhao}},
  \bibinfo{author}{\bibfnamefont{J.}~\bibnamefont{Peng}},
  \bibinfo{author}{\bibfnamefont{H.~Z.} \bibnamefont{Liang}},
  \bibinfo{author}{\bibfnamefont{P.}~\bibnamefont{Ring}}, \bibnamefont{and}
  \bibinfo{author}{\bibfnamefont{J.}~\bibnamefont{Meng}},
  \bibinfo{journal}{Phys. Rev. C} \textbf{\bibinfo{volume}{85}},
  \bibinfo{pages}{054310} (\bibinfo{year}{2012}).

\bibitem[{\citenamefont{Zhao et~al.}(2011{\natexlab{a}})\citenamefont{Zhao,
  Peng, Liang, Ring, and Meng}}]{Zhao2011Phys.Rev.Lett.122501}
\bibinfo{author}{\bibfnamefont{P.~W.} \bibnamefont{Zhao}},
  \bibinfo{author}{\bibfnamefont{J.}~\bibnamefont{Peng}},
  \bibinfo{author}{\bibfnamefont{H.~Z.} \bibnamefont{Liang}},
  \bibinfo{author}{\bibfnamefont{P.}~\bibnamefont{Ring}}, \bibnamefont{and}
  \bibinfo{author}{\bibfnamefont{J.}~\bibnamefont{Meng}},
  \bibinfo{journal}{Phys. Rev. Lett.} \textbf{\bibinfo{volume}{107}},
  \bibinfo{pages}{122501} (\bibinfo{year}{2011}{\natexlab{a}}).

\bibitem[{\citenamefont{Zhao et~al.}(2011{\natexlab{b}})\citenamefont{Zhao,
  Zhang, Peng, Liang, Ring, and Meng}}]{Zhao2011Phys.Lett.B181}
\bibinfo{author}{\bibfnamefont{P.~W.} \bibnamefont{Zhao}},
  \bibinfo{author}{\bibfnamefont{S.~Q.} \bibnamefont{Zhang}},
  \bibinfo{author}{\bibfnamefont{J.}~\bibnamefont{Peng}},
  \bibinfo{author}{\bibfnamefont{H.~Z.} \bibnamefont{Liang}},
  \bibinfo{author}{\bibfnamefont{P.}~\bibnamefont{Ring}}, \bibnamefont{and}
  \bibinfo{author}{\bibfnamefont{J.}~\bibnamefont{Meng}},
  \bibinfo{journal}{Phys. Lett. B} \textbf{\bibinfo{volume}{699}},
  \bibinfo{pages}{181} (\bibinfo{year}{2011}{\natexlab{b}}).

\bibitem[{\citenamefont{Hirata et~al.}(1997)\citenamefont{Hirata, Sumiyoshi,
  Tanihata, Sugahara, Tachibana, and Toki}}]{Hirata1997Nucl.Phys.A438}
\bibinfo{author}{\bibfnamefont{D.}~\bibnamefont{Hirata}},
  \bibinfo{author}{\bibfnamefont{K.}~\bibnamefont{Sumiyoshi}},
  \bibinfo{author}{\bibfnamefont{I.}~\bibnamefont{Tanihata}},
  \bibinfo{author}{\bibfnamefont{Y.}~\bibnamefont{Sugahara}},
  \bibinfo{author}{\bibfnamefont{T.}~\bibnamefont{Tachibana}},
  \bibnamefont{and} \bibinfo{author}{\bibfnamefont{H.}~\bibnamefont{Toki}},
  \bibinfo{journal}{Nucl. Phys. A} \textbf{\bibinfo{volume}{616}},
  \bibinfo{pages}{438 } (\bibinfo{year}{1997}).

\bibitem[{\citenamefont{Lalazissis et~al.}(1999)\citenamefont{Lalazissis,
  Raman, and Ring}}]{Lalazissis1999At.DataNucl.DataTables1}
\bibinfo{author}{\bibfnamefont{G.~A.} \bibnamefont{Lalazissis}},
  \bibinfo{author}{\bibfnamefont{S.}~\bibnamefont{Raman}}, \bibnamefont{and}
  \bibinfo{author}{\bibfnamefont{P.}~\bibnamefont{Ring}}, \bibinfo{journal}{At.
  Data Nucl. Data Tables} \textbf{\bibinfo{volume}{71}}, \bibinfo{pages}{1}
  (\bibinfo{year}{1999}).

\bibitem[{\citenamefont{Geng et~al.}(2005)\citenamefont{Geng, Toki, and
  Meng}}]{Geng2005Prog.Theor.Phys.785}
\bibinfo{author}{\bibfnamefont{L.~S.} \bibnamefont{Geng}},
  \bibinfo{author}{\bibfnamefont{H.}~\bibnamefont{Toki}}, \bibnamefont{and}
  \bibinfo{author}{\bibfnamefont{J.}~\bibnamefont{Meng}},
  \bibinfo{journal}{Prog. Theor. Phys.} \textbf{\bibinfo{volume}{113}},
  \bibinfo{pages}{785} (\bibinfo{year}{2005}).

\bibitem[{\citenamefont{Sun et~al.}(2008)\citenamefont{Sun, Montes, Geng,
  Geissel, Litvinov, and Meng}}]{Sun2008Phys.Rev.C025806}
\bibinfo{author}{\bibfnamefont{B.}~\bibnamefont{Sun}},
  \bibinfo{author}{\bibfnamefont{F.}~\bibnamefont{Montes}},
  \bibinfo{author}{\bibfnamefont{L.~S.} \bibnamefont{Geng}},
  \bibinfo{author}{\bibfnamefont{H.}~\bibnamefont{Geissel}},
  \bibinfo{author}{\bibfnamefont{Y.~A.} \bibnamefont{Litvinov}},
  \bibnamefont{and} \bibinfo{author}{\bibfnamefont{J.}~\bibnamefont{Meng}},
  \bibinfo{journal}{Phys. Rev. C} \textbf{\bibinfo{volume}{78}},
  \bibinfo{pages}{025806} (\bibinfo{year}{2008}).

\bibitem[{\citenamefont{Niu et~al.}(2009)\citenamefont{Niu, Sun, and
  Meng}}]{Niu2009Phys.Rev.C065806}
\bibinfo{author}{\bibfnamefont{Z.~M.} \bibnamefont{Niu}},
  \bibinfo{author}{\bibfnamefont{B.~H.} \bibnamefont{Sun}}, \bibnamefont{and}
  \bibinfo{author}{\bibfnamefont{J.}~\bibnamefont{Meng}},
  \bibinfo{journal}{Phys. Rev. C} \textbf{\bibinfo{volume}{80}},
  \bibinfo{pages}{065806} (\bibinfo{year}{2009}).

\bibitem[{\citenamefont{Li et~al.}(2012{\natexlab{a}})\citenamefont{Li, Niu,
  and Sun}}]{Li2012ActaPhys.Sin72601}
\bibinfo{author}{\bibfnamefont{Z.}~\bibnamefont{Li}},
  \bibinfo{author}{\bibfnamefont{Z.~M.} \bibnamefont{Niu}}, \bibnamefont{and}
  \bibinfo{author}{\bibfnamefont{B.}~\bibnamefont{Sun}}, \bibinfo{journal}{Acta
  Phys. Sin.} \textbf{\bibinfo{volume}{61}}, \bibinfo{pages}{072601}
  (\bibinfo{year}{2012}{\natexlab{a}}).

\bibitem[{\citenamefont{Xu et~al.}(2012)\citenamefont{Xu, Sun, Niu, Li, Qian,
  and Meng}}]{Xu2012}
\bibinfo{author}{\bibfnamefont{X. D.}~\bibnamefont{Xu}},
  \bibinfo{author}{\bibfnamefont{B.}~\bibnamefont{Sun}},
  \bibinfo{author}{\bibfnamefont{Z. M.}~\bibnamefont{Niu}},
  \bibinfo{author}{\bibfnamefont{Z.}~\bibnamefont{Li}},
  \bibinfo{author}{\bibfnamefont{Y.-Z.}~\bibnamefont{Qian}}, \bibnamefont{and}
  \bibinfo{author}{\bibfnamefont{J.}~\bibnamefont{Meng}}
  (\bibinfo{year}{2012}), \eprint{arXiv:1208.2341[nucl-th]}.

\bibitem[{\citenamefont{Zhao et~al.}(2010)\citenamefont{Zhao, Li, Yao, and
  Meng}}]{Zhao2010Phys.Rev.C54319}
\bibinfo{author}{\bibfnamefont{P.~W.} \bibnamefont{Zhao}},
  \bibinfo{author}{\bibfnamefont{Z.~P.} \bibnamefont{Li}},
  \bibinfo{author}{\bibfnamefont{J.~M.} \bibnamefont{Yao}}, \bibnamefont{and}
  \bibinfo{author}{\bibfnamefont{J.}~\bibnamefont{Meng}},
  \bibinfo{journal}{Phys. Rev. C} \textbf{\bibinfo{volume}{82}},
  \bibinfo{pages}{054319} (\bibinfo{year}{2010}).

\bibitem[{\citenamefont{Sun et~al.}(2011)\citenamefont{Sun, Zhao, and
  Meng}}]{Sun2011Sci.ChinaSer.G-Phys.Mech.Astron.210}
\bibinfo{author}{\bibfnamefont{B.}~\bibnamefont{Sun}},
  \bibinfo{author}{\bibfnamefont{P.}~\bibnamefont{Zhao}}, \bibnamefont{and}
  \bibinfo{author}{\bibfnamefont{J.}~\bibnamefont{Meng}},
  \bibinfo{journal}{Sci. China: Phys. Mech. Astron.} \textbf{\bibinfo{volume}{54}}, \bibinfo{pages}{210} (\bibinfo{year}{2011}),
  ISSN \bibinfo{issn}{1674-7348}.

\bibitem[{\citenamefont{Lu et~al.}(2012{\natexlab{b}})\citenamefont{Lu, Zhao,
  and Zhou}}]{Lu2012Phys.Rev.C11301}
\bibinfo{author}{\bibfnamefont{B.-N.} \bibnamefont{Lu}},
  \bibinfo{author}{\bibfnamefont{E.-G.} \bibnamefont{Zhao}}, \bibnamefont{and}
  \bibinfo{author}{\bibfnamefont{S.-G.} \bibnamefont{Zhou}},
  \bibinfo{journal}{Phys. Rev. C} \textbf{\bibinfo{volume}{85}},
  \bibinfo{pages}{011301} (\bibinfo{year}{2012}{\natexlab{b}}).

\bibitem[{\citenamefont{Chen
  et~al.}(2012{\natexlab{a}})}]{Chen2012Nucl.Phys.A71}
\bibinfo{author}{\bibfnamefont{L.}~\bibnamefont{Chen}} \bibnamefont{\textit{et~al}}.,
  \bibinfo{journal}{Nucl. Phys. A} \textbf{\bibinfo{volume}{882}},
  \bibinfo{pages}{71 } (\bibinfo{year}{2012}{\natexlab{a}}).

\bibitem[{\citenamefont{Nikolaus et~al.}(1992)\citenamefont{Nikolaus, Hoch, and
  Madland}}]{Nikolaus1992Phys.Rev.C1757}
\bibinfo{author}{\bibfnamefont{B.~A.} \bibnamefont{Nikolaus}},
  \bibinfo{author}{\bibfnamefont{T.}~\bibnamefont{Hoch}}, \bibnamefont{and}
  \bibinfo{author}{\bibfnamefont{D.~G.} \bibnamefont{Madland}},
  \bibinfo{journal}{Phys. Rev. C} \textbf{\bibinfo{volume}{46}},
  \bibinfo{pages}{1757} (\bibinfo{year}{1992}).

\bibitem[{\citenamefont{B\"urvenich et~al.}(2002)\citenamefont{B\"urvenich,
  Madland, Maruhn, and Reinhard}}]{Burvenich2002Phys.Rev.C44308}
\bibinfo{author}{\bibfnamefont{T.}~\bibnamefont{B\"urvenich}},
  \bibinfo{author}{\bibfnamefont{D.~G.} \bibnamefont{Madland}},
  \bibinfo{author}{\bibfnamefont{J.~A.} \bibnamefont{Maruhn}},
  \bibnamefont{and} \bibinfo{author}{\bibfnamefont{P.-G.}
  \bibnamefont{Reinhard}}, \bibinfo{journal}{Phys. Rev. C}
  \textbf{\bibinfo{volume}{65}}, \bibinfo{pages}{044308}
  (\bibinfo{year}{2002}).

\bibitem[{\citenamefont{Krieger et~al.}(1990)\citenamefont{Krieger, Bonche,
  Flocard, Quentin, and Weiss}}]{Krieger1990Nucl.Phys.A275}
\bibinfo{author}{\bibfnamefont{S.~J.} \bibnamefont{Krieger}},
  \bibinfo{author}{\bibfnamefont{P.}~\bibnamefont{Bonche}},
  \bibinfo{author}{\bibfnamefont{H.}~\bibnamefont{Flocard}},
  \bibinfo{author}{\bibfnamefont{P.}~\bibnamefont{Quentin}}, \bibnamefont{and}
  \bibinfo{author}{\bibfnamefont{M.~S.} \bibnamefont{Weiss}},
  \bibinfo{journal}{Nucl. Phys. A} \textbf{\bibinfo{volume}{517}},
  \bibinfo{pages}{275 } (\bibinfo{year}{1990}).

\bibitem[{\citenamefont{Bender et~al.}(2000{\natexlab{a}})\citenamefont{Bender,
  Rutz, Reinhard, and Maruhn}}]{Bender2000Eur.Phys.J.A59}
\bibinfo{author}{\bibfnamefont{M.}~\bibnamefont{Bender}},
  \bibinfo{author}{\bibfnamefont{K.}~\bibnamefont{Rutz}},
  \bibinfo{author}{\bibfnamefont{P.-G.} \bibnamefont{Reinhard}},
  \bibnamefont{and} \bibinfo{author}{\bibfnamefont{J.~A.}
  \bibnamefont{Maruhn}}, \bibinfo{journal}{Eur. Phys. J. A}
  \textbf{\bibinfo{volume}{8}}, \bibinfo{pages}{59}
  (\bibinfo{year}{2000}{\natexlab{a}}).

\bibitem[{\citenamefont{Bender et~al.}(2000{\natexlab{b}})\citenamefont{Bender,
  Rutz, Reinhard, and Maruhn}}]{Bender2000Eur.Phys.J.A467}
\bibinfo{author}{\bibfnamefont{M.}~\bibnamefont{Bender}},
  \bibinfo{author}{\bibfnamefont{K.}~\bibnamefont{Rutz}},
  \bibinfo{author}{\bibfnamefont{P.-G.} \bibnamefont{Reinhard}},
  \bibnamefont{and} \bibinfo{author}{\bibfnamefont{J.~A.}
  \bibnamefont{Maruhn}}, \bibinfo{journal}{Eur. Phys. J. A}
  \textbf{\bibinfo{volume}{7}}, \bibinfo{pages}{467}
  (\bibinfo{year}{2000}{\natexlab{b}}).

\bibitem[{\citenamefont{Zhao et~al.}(2009)\citenamefont{Zhao, Sun, and
  Meng}}]{Zhao2009Chin.Phys.Lett.112102}
\bibinfo{author}{\bibfnamefont{P.}~\bibnamefont{Zhao}},
  \bibinfo{author}{\bibfnamefont{B.}~\bibnamefont{Sun}}, \bibnamefont{and}
  \bibinfo{author}{\bibfnamefont{J.}~\bibnamefont{Meng}},
  \bibinfo{journal}{Chin. Phys. Lett.} \textbf{\bibinfo{volume}{26}},
  \bibinfo{pages}{112102} (\bibinfo{year}{2009}).

\bibitem[{\citenamefont{Inglis}(1956)}]{Inglis1956Phys.Rev.1786}
\bibinfo{author}{\bibfnamefont{D.}~\bibnamefont{Inglis}},
  \bibinfo{journal}{Phys. Rev.} \textbf{\bibinfo{volume}{103}},
  \bibinfo{pages}{1786} (\bibinfo{year}{1956}).

\bibitem[{\citenamefont{Belyaev}(1961)}]{Belyaev1961Nucl.Phys.A322}
\bibinfo{author}{\bibfnamefont{S.}~\bibnamefont{Belyaev}},
  \bibinfo{journal}{Nucl. Phys. A} \textbf{\bibinfo{volume}{24}},
  \bibinfo{pages}{322} (\bibinfo{year}{1961}).

\bibitem[{\citenamefont{Volkov}(1972)}]{Volkov1972Phys.Lett.B1}
\bibinfo{author}{\bibfnamefont{A.~B.} \bibnamefont{Volkov}},
  \bibinfo{journal}{Phys. Lett. B} \textbf{\bibinfo{volume}{41}},
  \bibinfo{pages}{1} (\bibinfo{year}{1972}), ISSN \bibinfo{issn}{0370-2693}.

\bibitem[{\citenamefont{Ring et~al.}(1997)\citenamefont{Ring, Gambhir, and
  Lalazissis}}]{Ring1997Comput.Phys.Commun.77}
\bibinfo{author}{\bibfnamefont{P.}~\bibnamefont{Ring}},
  \bibinfo{author}{\bibfnamefont{Y.~K.} \bibnamefont{Gambhir}},
  \bibnamefont{and} \bibinfo{author}{\bibfnamefont{G.~A.}
  \bibnamefont{Lalazissis}}, \bibinfo{journal}{Comput. Phys. Commun.}
  \textbf{\bibinfo{volume}{105}}, \bibinfo{pages}{77 } (\bibinfo{year}{1997}).

\bibitem[{\citenamefont{Audi et~al.}(2003)\citenamefont{Audi, Wapstra, and
  Thibault}}]{Audi2003Nucl.Phys.A337}
\bibinfo{author}{\bibfnamefont{G.}~\bibnamefont{Audi}},
  \bibinfo{author}{\bibfnamefont{A.~H.} \bibnamefont{Wapstra}},
  \bibnamefont{and} \bibinfo{author}{\bibfnamefont{C.}~\bibnamefont{Thibault}},
  \bibinfo{journal}{Nucl. Phys. A} \textbf{\bibinfo{volume}{729}},
  \bibinfo{pages}{337 } (\bibinfo{year}{2003}).

\bibitem[{\citenamefont{Samyn et~al.}(2002)\citenamefont{Samyn, Goriely,
  Heenen, Pearson, and Tondeur}}]{Samyn2002Nucl.Phys.A142}
\bibinfo{author}{\bibfnamefont{M.}~\bibnamefont{Samyn}},
  \bibinfo{author}{\bibfnamefont{S.}~\bibnamefont{Goriely}},
  \bibinfo{author}{\bibfnamefont{P.}~\bibnamefont{Heenen}},
  \bibinfo{author}{\bibfnamefont{J.}~\bibnamefont{Pearson}}, \bibnamefont{and}
  \bibinfo{author}{\bibfnamefont{F.}~\bibnamefont{Tondeur}},
  \bibinfo{journal}{Nucl. Phys. A} \textbf{\bibinfo{volume}{700}},
  \bibinfo{pages}{142} (\bibinfo{year}{2002}).

\bibitem[{\citenamefont{Meng and Ring}(1996)}]{Meng1996Phys.Rev.Lett.3963}
\bibinfo{author}{\bibfnamefont{J.}~\bibnamefont{Meng}} \bibnamefont{and}
  \bibinfo{author}{\bibfnamefont{P.}~\bibnamefont{Ring}},
  \bibinfo{journal}{Phys. Rev. Lett.} \textbf{\bibinfo{volume}{77}},
  \bibinfo{pages}{3963} (\bibinfo{year}{1996}).

\bibitem[{\citenamefont{Meng}(1998)}]{Meng1998Nucl.Phys.A3}
\bibinfo{author}{\bibfnamefont{J.}~\bibnamefont{Meng}}, \bibinfo{journal}{Nucl.
  Phys. A} \textbf{\bibinfo{volume}{635}}, \bibinfo{pages}{3}
  (\bibinfo{year}{1998}).

\bibitem[{\citenamefont{Zhou et~al.}(2010)\citenamefont{Zhou, Meng, Ring, and
  Zhao}}]{Zhou2010Phys.Rev.C011301R}
\bibinfo{author}{\bibfnamefont{S.-G.} \bibnamefont{Zhou}},
  \bibinfo{author}{\bibfnamefont{J.}~\bibnamefont{Meng}},
  \bibinfo{author}{\bibfnamefont{P.}~\bibnamefont{Ring}}, \bibnamefont{and}
  \bibinfo{author}{\bibfnamefont{E.-G.} \bibnamefont{Zhao}},
  \bibinfo{journal}{Phys. Rev. C} \textbf{\bibinfo{volume}{82}},
  \bibinfo{pages}{011301(R)} (\bibinfo{year}{2010}).

\bibitem[{\citenamefont{Li et~al.}(2012{\natexlab{b}})\citenamefont{Li, Meng,
  Ring, Zhao, and Zhou}}]{Li2012Phys.Rev.C24312}
\bibinfo{author}{\bibfnamefont{L.}~\bibnamefont{Li}},
  \bibinfo{author}{\bibfnamefont{J.}~\bibnamefont{Meng}},
  \bibinfo{author}{\bibfnamefont{P.}~\bibnamefont{Ring}},
  \bibinfo{author}{\bibfnamefont{E.-G.}~\bibnamefont{Zhao}}, \bibnamefont{and}
  \bibinfo{author}{\bibfnamefont{S.-G.}~\bibnamefont{Zhou}},
  \bibinfo{journal}{Phys. Rev. C} \textbf{\bibinfo{volume}{85}},
  \bibinfo{pages}{024312} (\bibinfo{year}{2012}{\natexlab{b}}).

\bibitem[{\citenamefont{Chen et~al.}(2012{\natexlab{b}})\citenamefont{Chen, Li,
  Liang, and Meng}}]{Chen2012Phys.Rev.C67301}
\bibinfo{author}{\bibfnamefont{Y.}~\bibnamefont{Chen}},
  \bibinfo{author}{\bibfnamefont{L.}~\bibnamefont{Li}},
  \bibinfo{author}{\bibfnamefont{H.}~\bibnamefont{Liang}}, \bibnamefont{and}
  \bibinfo{author}{\bibfnamefont{J.}~\bibnamefont{Meng}},
  \bibinfo{journal}{Phys. Rev. C} \textbf{\bibinfo{volume}{85}},
  \bibinfo{pages}{067301} (\bibinfo{year}{2012}{\natexlab{b}}).

\bibitem[{\citenamefont{Koepf and Ring}(1990)}]{Koepf1990Nucl.Phys.A279}
\bibinfo{author}{\bibfnamefont{W.}~\bibnamefont{Koepf}} \bibnamefont{and}
  \bibinfo{author}{\bibfnamefont{P.}~\bibnamefont{Ring}},
  \bibinfo{journal}{Nucl. Phys. A} \textbf{\bibinfo{volume}{511}},
  \bibinfo{pages}{279 } (\bibinfo{year}{1990}).

\bibitem[{\citenamefont{Geng}(2005)}]{Geng2005}
\bibinfo{author}{\bibfnamefont{L.}~\bibnamefont{Geng}}, Ph.D. thesis,
  \bibinfo{school}{Osaka University}, \bibinfo{year}{2005} (unpublished).
\end{thebibliography}

\end{document}